\documentclass[%
 reprint,
superscriptaddress,
nofootinbib,
 amsmath,amssymb,
aps,
tightenlines
]{revtex4-2}

\usepackage{aas_macros}
\usepackage{graphicx}
\usepackage{dcolumn}
\usepackage{bm}
\usepackage{xcolor}
\definecolor{linkcolor}{rgb}{0.6,0,0}
\definecolor{citecolor}{rgb}{0,0,0.75}
\definecolor{urlcolor}{rgb}{0.12,0.46,0.7}
\usepackage[breaklinks, colorlinks, urlcolor=urlcolor, linkcolor=linkcolor,citecolor=citecolor,pdfencoding=auto]{hyperref}
\usepackage{amsmath}
\usepackage{ulem}
\usepackage{xspace}
\usepackage{booktabs}
\usepackage{xcolor,colortbl}
\usepackage{multirow}
\usepackage[autostyle=false, style=english]{csquotes}
\MakeOuterQuote{"}

\newcommand{\nver}{\hat{\mathbf{n}}}
\DeclareMathOperator*{\argmax}{argmax}

\begin{document}

\preprint{APS/123-QED}

\title{Inference of gravitational lensing and patchy reionization with future CMB data}

\author{Federico Bianchini}
\email{federico.bianxini@gmail.com}
\affiliation{Kavli Institute for Particle Astrophysics and Cosmology, Stanford University, 452 Lomita Mall, Stanford, 
CA, 94305, USA}
\affiliation{SLAC National Accelerator Laboratory, 2575 Sand Hill Road, Menlo Park, CA 94025}
\affiliation{Department of Physics, Stanford University, 382 Via Pueblo Mall, Stanford, CA 94305}

\author{Marius Millea}
\email{mariusmillea@gmail.com}
\affiliation{Department of Physics, University of California, Berkeley, CA 94720}
\affiliation{Department of Physics, University of California, Davis, CA 95616}

\date{\today}

\begin{abstract}
We develop an optimal Bayesian solution for jointly inferring secondary signals in the Cosmic Microwave Background (CMB) originating from gravitational lensing and from patchy screening during the epoch of reionization. This method is able to extract full information content from the data, improving upon previously considered quadratic estimators for lensing and screening. We forecast constraints using the Marginal Unbiased Score Expansion (MUSE) method, and show that they are largely dominated by CMB polarization, and depend on the exact details of reionization. For models consistent with current data which produce the largest screening signals, a detection (3\,$\sigma$) of the cross-correlation between lensing and screening is possible with SPT-3G, and a detection of the auto-correlation is possible with CMB-S4. Models with the lowest screening signals evade the sensitivity of SPT-3G, but are still possible to detect with CMB-S4 via their lensing cross-correlation.
\end{abstract}

\maketitle


\section{\label{sec:intro} Introduction}
The large-scale structure (LSS) of the universe is backlit by cosmic microwave background (CMB) photons as they travel from the last scattering surface towards us.
Maps of the CMB anisotropies can therefore be used to image gravitational potentials -- through weak gravitational lensing, integrated Sachs-Wolfe and Rees-Sciama effects -- and the gas distribution -- through Thomson and inverse Compton scattering processes like the Sunyaev-Zel'dovich effects \citep[e.g.,][]{sunyaev70,sunyaev72,carlstrom02,lewis06,aghanim08}. 
The current generation of CMB surveys such as \textit{Planck} \citep[][]{planck18-6}, SPT \citep{reichardt20} and ACT \citep{aiola20} has started to tap into the promising potential of these CMB secondary anisotropies. 
Next-generation experiments -- including the Simons Observatory (SO) \citep{SO}, FYST/CCAT-prime \citep{aravena19}, and CMB-S4 \citep{cmbs4collab19} -- will provide a transformative high fidelity view of the secondary CMB anisotropies in intensity and polarization over large areas of the sky, revealing fundamental insights into both cosmology and astrophysics \citep{chang22}.

Mapping out the spatial distribution of diffuse ionized gas throughout the universe can, for example, help us understand the physics of reionization, at high redshifts ($z \gtrsim 6$), and of the intergalactic medium (IGM), at lower redshifts ($z \lesssim 6$) \citep[e.g.,][]{gruzinov98,mortonson07}.
One approach to achieve this is by searching for the characteristic spatially-dependent suppression of the CMB temperature and polarization anisotropies produced by different scattering histories along different line of sights, an effect known as "patchy screening".
The magnitude of the effect is proportional to $e^{-\tau(\nver)}$, where $\tau(\nver)$ is the direction-dependent optical depth.

The so-called "quadratic estimator" (QE) has become the workhorse for extracting sources of statistical anisotropies, such as patchy screening and lensing, from CMB maps \citep[e.g.,][]{zaldarriaga99,hu02a,hu03}.
The QE in the context of inhomogeneous optical depth reconstruction has been introduced by \citet{dvorkin09} and applied to WMAP and \textit{Planck} data in \citep[][]{gluscevic13,namikawa18,namikawa21} but the spatial fluctuations of $\tau$ have not been detected yet.
While the QE has been successfully used on current datasets, it has some shortcomings.
First, the presence of other distorting fields, like lensing, point-sources, and inhomogeneous noise, will introduce additional non-Gaussianities in the data which in turn, lead to biases in the reconstructed field \citep{su11}.
"Bias-hardened" estimators offer a solution to this problem at the cost of a signal-to-noise ($S/N$) degradation, which can be as large as $\approx 40\%$ \citep[e.g.,][]{su11,namikawa21}. 
Second, the QE will become significantly sub-optimal at the instrumental noise levels soon-to-be reached by the most sensitive experiments \citep[][]{PB19,millea21}.
At these depths, secondary anisotropies, rather than instrumental noise, limit the variance of the estimated field. 
To improve upon the QE, a variety of methods based on the full CMB Bayesian posterior have been proposed to extract the higher-order information and restore near-optimality \citep[e.g.,][]{hirata03a,hirata03b,carron17,millea19,millea20,millea22}.
Machine-learning approaches are also being currently investigated but while promising, additional work towards the characterization of these methods and their biases is needed before they can be reliably applied to real data \citep{caldeira18,guzman21}.

In this paper, we develop a complete Bayesian solution that unifies the optimal inference of the optical depth $\tau$ and CMB lensing potential $\phi$ together with delensing and cosmological parameter inference.
Our method presents a number of appealing features.
First of all, by making use of the full Bayesian posterior, the method is capable of optimally extracting the information content in CMB data at all noise levels. 
Furthermore, by simultaneously forward modeling the effects of lensing and screening on CMB observables, we are able to naturally account for any contamination (of lensing to screening and viceversa) in the reconstruction and to enhance the sensitivity to screening through the cosmic variance reduction due to delensing.
Our method can also internally measure the correlation $\langle\phi\tau\rangle$ between the CMB lensing potential and optical depth fluctuations, which is larger than $\langle\tau\tau\rangle$, contains additional information on relation between the ionized gas and dark matter distribution, and is expected to be detected with upcoming CMB surveys \cite{feng19}. 
Finally, the CMB data are effectively "descreened" by our procedure, which can in turn mitigate any residual $B$-mode bias from screening to tensor-to-scalar ratio $r$ searches (although the contamination is expected to be at the level of $r\sim 10^{-4}$ for standard reionization histories, see, e.g., \citep{roy20,mukherjee19}, therefore not a significant concern for experiments targeting $\mathcal{O}(10^{-3})$).

We begin the paper with a review of the theoretical background and the effect of patchy screening on CMB observables in Sec.~\ref{sec:theory}.
We present our method and illustrate its performance in Sec.~\ref{sec:results}, before concluding in Sec.~\ref{sec:conclusions}.

\section{\label{sec:theory} Modeling}
\subsection{Optical depth}
The electron scattering optical depth measures the integrated electron density along the line-of-sight and is given by
\begin{equation} 
    \tau(\nver)=\sigma_{\rm{T}} \int {\rm d} \chi \, a\, n_e(\nver, \chi),
\end{equation}
where $n_{\rm{e}}(\nver, \chi)$ is the free electron number density at comoving distance $\chi$ along the direction $\nver$, $\sigma_T$ is the Thomson scattering cross-section, and $a$ is the scale factor.
The mean number density of free electrons can be expressed as $\bar{n}_e= (1-3/4Y_P)\bar{\rho}_{b0}/m_pa^{-3}\bar{x}_e=\bar{n}_{p0}\bar{x}_e/a^3$, where $Y_P$ is the primordial helium abundance, $\rho_{b0}$ is the present-day baryon density, the proton mass is $m_p$, $\bar{x}_e$ is the mean ionization fraction, and we assumed that helium is singly ionized.
We model the evolution of the volume-averaged ionization fraction $\bar{x}_e$ using a simple tanh fitting function parametrized by  the redshift of reionization $z_{\rm re}$, defined as the redshift at which $\bar{x}_e$ is half of its maximum, and the duration of reionization $\Delta_z$, i.e. the difference between the redshifts at which the universe is 5\% and 95\% reionized \citep{lewis08} 
\begin{equation}
\label{eq:xe}
    \bar{x}_e(z)=\frac{1}{2}\left[1+\tanh \left(\frac{y_{\mathrm{re}}-y}{\Delta y_{\mathrm{re}}}\right)\right],
\end{equation}
where $y(z)=(1+z)^{3/2}$, $y_{\rm re}=y(z_{\rm re})$, and $\Delta y_{\rm re} = \sqrt{1+z_{\rm re}}\Delta_z$. 

Perturbations in the free electron number density $\delta n_e$ can be sourced both by ionization fluctuations $\delta_x =\delta x_e/\bar{x}_e$ and by inhomogeneities in the gas density $\delta$ \citep[][]{ mortonson07}. 
The former are only generated during the epoch of reionization while the latter are produced both when reionization occurs as well as in the post-reionization universe.
Fluctuations in the free electron density will then induce spatial fluctuations in the optical depth \citep[e.g.,][]{holder06}:
\begin{align}
    \tau(\nver)&=\sigma_{\rm{T}} n_{p0} \int \frac{{\rm d}\chi}{a^2}  \bar{x}_e(\chi)\left(1 + \delta_x(\nver, \chi)\right) \left(1+\delta(\nver, \chi)\right)\\
    &= \bar{\tau} + \delta\tau(\nver),
\end{align} 
The contribution to the $\tau$ anisotropies from the spatial distribution of free electrons in galaxies and clusters is then given by the redshifts below which $\bar{x}_e = 1$.

Under the Limber approximation \citep{limber53}, valid for angular scales $\ell\gg 20$ relevant for this work, the angular power spectrum of the optical depth fluctuations can be evaluated as 
\begin{equation}
    C_\ell^{\tau \tau}=\sigma_{T}^{2} n_{p 0}^{2}\int \frac{{\rm d}\chi}{a^{4} \chi^{2}} P_{\delta_ e\delta_e}\left(k_\ell, \chi\right),
\end{equation}
where $ P_{\delta_e\delta_e}$ is the power spectrum of the  density-weighted ionization fraction fluctuations, and $k_\ell = (\ell+\frac{1}{2})/\chi$. 

The quantity $P_{\delta_e\delta_e}$ encodes all the relevant astrophysical aspects of the EoR, including its morphology and timing. 
Modeling reionization and the spatial distribution of free electrons is a challenging task due to the complexity of the physical processes involved and the limited observational access we currently have to those cosmic epochs.
Therefore, instead of specifying a given physical mechanism of reionization \citep[e.g.,][]{furlanetto04}, we choose to adopt the "bubble model" introduced in \citep{wang06,mortonson07,dvorkin09} that allows us to phenomenologically parametrize the \textsc{Hii} spectrum,  $P_{\delta_e \delta_e}$.
In this framework, the \textsc{Hii} regions around the ionizing sources, such as galaxies or quasars, are biased tracers of the underlying dark matter halos.
The ionized bubbles are assumed to be spherical with an average radius $\bar{R}$ (in Mpc) while their radii distribution is modeled as a log-normal distribution of width $\sigma_{\ln\!R}$, i.e. skewed towards smaller bubble sizes.
The size and evolution of these \textsc{Hii} bubbles are sensitive to the mass and brightness of the ionizing sources.
As time advances, the \textsc{Hii} bubbles grow in size and percolate, eventually leading to a complete reionization of the intergalactic medium.
In App.~\ref{app:halo_model} we provide a more detailed discussion of the halo model.

To summarize, our inhomogenous reionization modeling is described by a set of four parameters: the redshift and duration of reionization $\{z_{\rm re},\Delta_z\}$, which specify the mean ionization history $\bar{x}_e(z)$, and the characteristic size and standard deviation of the log-normal bubble radius distribution, $\{\bar{R}, \sigma_{\ln\!R}\}$.

\subsection{Cross-correlation with CMB lensing}
In this work we are also interested in evaluating the correlation between the optical depth fluctuations and the integrated matter distribution along the line-of-sight as traced by the CMB lensing potential $\phi(\nver)$ defined as \citep[e.g.,][]{lewis06}
\begin{equation}
\label{eq:phi}
    \phi(\nver)= -2\int {\rm d} \chi\, \frac{\chi-\chi_*}{\chi\chi_*} \Psi(\nver,\chi).
\end{equation}
Here, $\Psi(\nver,\chi)$ is the Weyl (gravitational) potential, that in standard cosmologies can be directly related to the comoving matter perturbations $\delta(\nver,\chi)$ through the Poisson equation, and $\chi_*$ is the comoving distance to the last scattering surface at $z_* \simeq 1090$.

A cross-correlation between optical depth fluctuations and CMB lensing is naturally expected since the same dark matter halos signposted by the \textsc{Hii} bubbles at high redshift ($z \gtrsim 6$) and by galaxies and galaxy clusters at low redshift ($z \lesssim 6$) act as lenses for the CMB photons.
In fact, the CMB lensing kernel has non-zero support out to high redshift and it is about half of its maximum around $z\approx 10$.

Using Eqn.~\eqref{eq:phi}, the cross-power spectrum between the fluctuating part of $\tau$ and CMB lensing potential takes the following form:
\begin{equation}
    C_\ell^{\phi \tau}=\frac{3 H_{0}^{2} \Omega_{m} \sigma_{T} n_{p 0}}{c \ell^{2}} \int \frac{{\rm d} \chi}{a^{3}} \frac{\chi-\chi_*}{\chi\chi_*} P_{\delta \delta_{e}}(k_\ell, \chi),
\end{equation}
where $H_0$ is the Hubble constant and $P_{\delta \delta_{e}}(k_\ell, \chi)$ denotes the three-dimensional cross-spectrum between the matter density contrast $\delta$ and the free electron fluctuations $\delta_e$. 

\subsection{Fiducial scenarios}
While we do not specify a given physical model for reionization, here we consider two limiting reionization scenarios to study the performance of our approach and to get a sense for the expected $S/N$ with current and upcoming surveys.

In the first case (optimistic model), reionization starts early on, proceeds for an extended period of time, and is driven by larger bubbles with a broad radii distribution spread. 
For this scenario we fix $\{z_{\rm re},\Delta_z, \bar{R}, \sigma_{\ln\!R}\}=\{10,4,5,\ln(2)\}$. Aside from giving a signal which is large but not currently detectable, a value of $\sigma_{\ln\!R}\,{=}\,\ln(2)$, is on the upper end of allowed values given existing cross-correlations between QE reconstructions of $\tau$ and Compton $y$-maps using {\it Planck} data \cite{namikawa21}.
In the second case (pessimistic model), reionization occurs at lower redshift and its duration is shorter, while the bubbles that reionize the universe are smaller in size and their radii distribution is narrower. 
When analyzing this scenario we set $\{z_{\rm re},\Delta_z, \bar{R}, \sigma_{\ln\!R}\}=\{7,1,1,0.1\}$. As we will see, the nomenclature of optimistic or pessimistic refers to the prospects of detecting the patchy screening effect in the two cases. 

\begin{figure*}
\includegraphics[width=\textwidth]{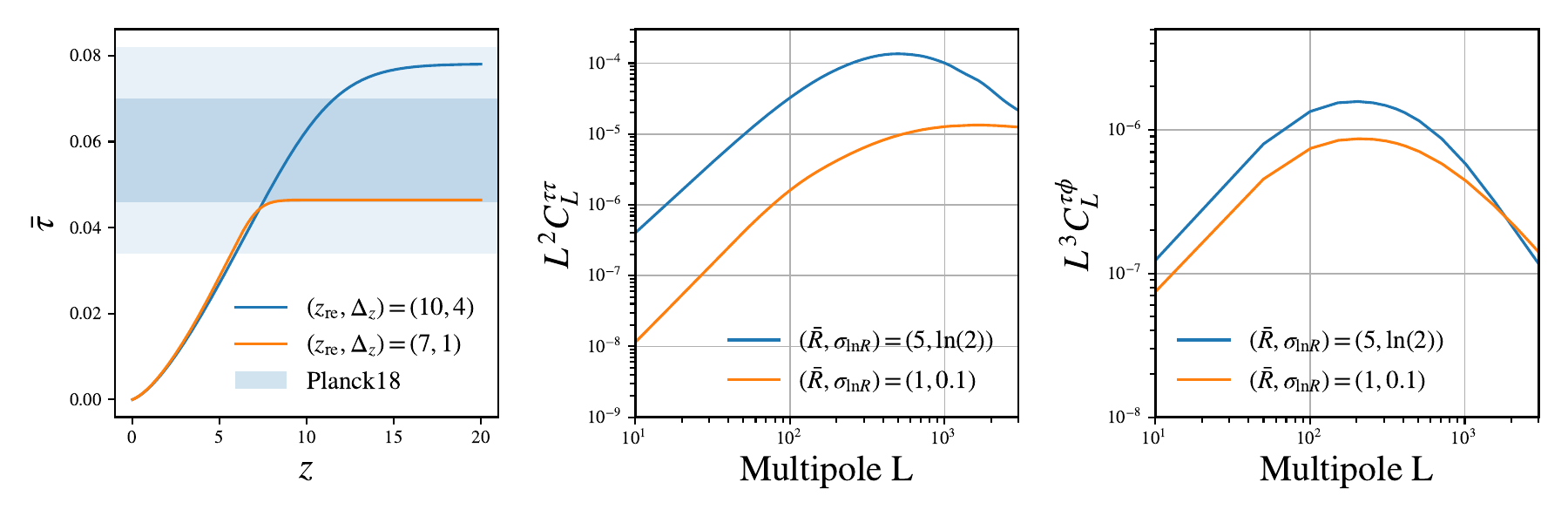}
\caption{\textit{Left:} Redshift evolution of the electron scattering optical depth based on the ionization fraction model of Eqn.~\eqref{eq:xe} assuming two different reionization histories (blue and orange solid lines). The shaded blue bands show the 1 and 2$\sigma$ constraints on $\bar{\tau}$ from \textit{Planck} \citep{planck18-6}.   \textit{Middle:} Angular power spectrum of the optical depth calculated from the bubble model. The colors refer to the reionization history parameters of the left panel. \textit{Right}: Cross-power spectrum between optical depth and CMB lensing potential (same color coding as the middle panel).}
\label{fig:fid_spectra}
\end{figure*}

In the left hand panel of Fig.~\ref{fig:fid_spectra}, we show the two reionization histories.
The higher redshift of reionization in the optimistic model results in a larger integrated mean optical depth $\bar{\tau}$ than in the pessimistic model. 
Both values are consistent within $2\sigma$ with the constraint from \textit{Planck} on $\bar{\tau} = 0.058 \pm 0.012$ \citep[][]{planck18-6}.
The middle and right panels of Fig.~\ref{fig:fid_spectra} show the optical depth fluctuation angular auto spectrum and its cross-correlation with CMB lensing, respectively.
Qualitatively, a longer duration of reionization translates to more pronounced fluctuations in $\tau$, corresponding to a larger overall amplitude of $C_{\ell}^{\tau\tau}$ and $C_{\ell}^{\phi\tau}$. 
The peak of these spectra is mostly determined by the effective radius of the \textsc{Hii} bubbles, with larger bubbles enhancing the power spectra at lower multipoles.
 
\subsection{Hierarchical model for CMB data}
The patchy nature of the reionization process at $z \gtrsim 6$ and the clumpiness of the spatial distribution of free electrons at lower redshifts ($z \lesssim 6$) lead to an anisotropic optical depth, $\tau = \tau(\nver)$. 
In turn, spatial variations of the optical depth will introduce new observational signatures in the CMB sky. 

Three main effects can be identified. 
First, Thomson scattering of the remote CMB temperature quadrupole by ionized bubbles generates new polarization \citep[e.g.,][]{gruzinov98,mortonson07,dore07,dvorkin09b,battaglia13,paul20}.\footnote{While new $E$-mode polarization is generated even in the case of homogeneous reionization, additional $B$-modes can only be sourced if the distribution of free electrons is inhomogeneous.}
Second, the radial peculiar velocity of ionized bubbles relative to the observer sources a temperature fluctuation through the kinematic Sunyaev-Zel'dovich (kSZ) effect \citep[e.g.,][]{santos03,zahn05,battaglia13,gorce20,choudhury20,paul20}.
Third, anisotropies in the optical depth will induce a spatially-dependent screening of the primary CMB anisotropies caused by the CMB photons being scattered into and out of our line-of-sight \citep[e.g.,][]{dvorkin09b, natarajan13}.
In this paper, we focus exclusively on the latter effect.

Considering an unperturbed CMB field $f \in \{T,Q,U\}$, the map-level effect of anisotropic screening is to spatially modulate the amplitude of the CMB temperature and polarization anisotropies by a factor $e^{-\tau(\nver)}$ as
\begin{equation}
    (\mathbb{S}(\tau)f)(\nver) = e^{-\tau(\nver)}f(\nver),
\end{equation}
where we have introduced the screening operator $\mathbb{S}(\tau)$.

On the other hand, the gravitational potentials associated to the intervening matter distribution between us and the last scattering surface induce a remapping of the primary CMB anisotropies, which we model (in the absence of screening) as 
\begin{equation}
    (\mathbb{L}(\phi)f)(\nver) = f(\nver + \nabla\phi(\nver)),
\end{equation}
where we have similarly introduced the lensing operator $\mathbb{L}(\phi)$.

In our treatment of the two effects, we make the simplifying assumption that patchy screening happens first and the lensing operator is then applied to the screened CMB fields:
\begin{equation}
(\mathbb{L}(\phi) \, \mathbb{S}(\tau) f)(\nver) = e^{-\tau(\nver+ \nabla\phi(\nver))}f(\nver + \nabla\phi(\nver)).
\end{equation}

Given these operators, we can model observed CMB data as
\begin{equation}
    d=\mathbb{A} \, \mathbb{L}(\phi) \, \mathbb{S}(\tau) f+n,
\end{equation}
where $d$ is the data, $f$ are the unlensed and unscreened CMB fields, and $n$ is the instrumental noise. 
The linear operator $\mathbb{A}\equiv \mathbb{M}\,\mathbb{T}\,\mathbb{B}$ encodes the effects of the instrumental beam $\mathbb{B}$, the transfer function $\mathbb{T}$, and any pixel mask $\mathbb{M}$. 

The goal of this analysis is to simultaneously extract lensing and patchy screening effects from noisy CMB data. 
There are different ways to cast this problem. 
In this work we choose to introduce three amplitude parameters $\theta=\left\{ A_{\phi\phi}, A_{\tau\tau}, A_{\phi\tau}\right\}$ that rescale some fiducial CMB lensing and optical depth fiducial isotropic power spectra as $C_\ell^{XY} = A_{XY}C_{\ell}^{XY,\rm fid}$ with $X,Y\in \{\phi,\tau\}$. 
This allows us to rapidly explore the reionization parameter space and to efficiently forecast the detection capabilities of various surveys in different scenarios.
While we adopt this approach for the scope of this paper, we note that nothing prevents us from carrying out the statistical inference directly at the power spectrum level \citep[e.g.,][]{millea22}, or on the phenomenological parameters $(\bar R, \sigma_{\ln\!R})$ directly.

With this definition, the CMB screening and lensing problem can then be formulated as
\begin{align}
\label{eq:fw_model}
\left( A_{\phi\phi}, A_{\tau\tau}, A_{\phi\tau}\right) & \sim \mathcal{U}\left(0,\infty\right) \\
f & \sim \mathcal{N}\left(0, \mathbb{C}_f\right) \\
\phi,\tau & \sim \mathcal{N}\left(0, \mathbb{C}_{\phi\tau}(\theta)\right) \\
d & \sim \mathcal{N}\left(\mathbb{L}(\phi) \mathbb{S}(\tau) f,\mathbb{C}_{n}\right),
\end{align}
where $\mathcal{U}$ and $\mathcal{N}$ denote uniform and Gaussian distributions, respectively.
We assume that both the unperturbed CMB and the experimental noise are Gaussian random fields described by covariance operators $\mathbb{C}_n$ and $\mathbb{C}_f$, respectively, where the latter reads
\begin{align}
\mathbb{C}_f & = \left[\begin{array}{ccc}
C_{\ell}^{T T} & C_{\ell}^{T E} & 0 \\
C_{\ell}^{T E} & C_{\ell}^{E E} & 0 \\
0 & 0 & C_{\ell}^{B B}
\end{array}\right].
\end{align}
Here we adopt the prior that both $\phi$ and $\tau$ are correlated Gaussian random fields with covariance given by
\begin{align}
\mathbb{C}_{\phi\tau}\left(\theta\right) & =  \left[\begin{array}{ll}
A_{\phi\phi} C_{\ell}^{\phi \phi} & A_{\phi\tau} C_{\ell}^{\phi \tau} \\
A_{\phi\tau} C_{\ell}^{\phi \tau} & A_{\tau\tau} C_{\ell}^{\tau \tau}
\end{array}\right].
\end{align}
where we emphasize the explicit dependence of $\mathbb{C}_{\phi\tau}$ on the spectral amplitude parameters $\theta$. Although the $\tau$ field is not expected to be Gaussian due to the non-linear growth of the ionizing bubbles, our choice to take a Gaussian prior is made for simplicity and because it is in some the sense the conservative choice for a first detection. In this way, inference will not attempt to extract any information from higher-order statistics, which are more difficult to realistically model.

With these ingredients, the joint likelihood distribution takes the following form
\begin{widetext}
\begin{align}
\label{eq:jointposterior}
    \mathcal{P}(d, \underbrace{f, \phi, \tau}_{z} \mid \underbrace{A_{\tau\tau}, A_{\phi\tau},A_{\phi\phi}}_{\theta}) \propto \frac{\exp \left\{-\frac{(d-\mathbb{A}\,\mathbb{L}(\phi) \, \mathbb{S}(\tau) f)^{2}}{2 \mathbb{C}_{n}}\right\}}{\operatorname{det} \mathbb{C}_{n}^{1 / 2}} \frac{\exp \left\{-\frac{f^{2}}{2 \mathbb{C}_{f}}\right\}}{\operatorname{det} \mathbb{C}_{f}^{1 / 2}} \frac{\exp \left\{-\frac{(\phi\,\oplus\,\tau)^{2}}{2 \mathbb{C}_{\phi\tau}\left(\theta\right)}\right\}}{\operatorname{det} \mathbb{C}_{\phi\tau}\left(\theta\right)^{1 / 2}},
\end{align}
\end{widetext}
where we have explicitly highlighted that the latent space $z$, over which marginalization will be performed, is composed by $\{f,\phi,\tau\}$, and we recall that the parameters of interest $\theta$ are the power spectra amplitudes $\left\{ A_{\phi\phi}, A_{\tau\tau}, A_{\phi\tau}\right\}$.
In Eqn.~\eqref{eq:jointposterior}, we denote the concatenation of the two abstract vectors $\phi$ and $\tau$ using the symbol $\oplus$, and we use the shorthand $x^2/\mathbb{N}\equiv x^\dagger \mathbb{N}^{-1}x$.

We give the posterior function in Eqn.~\eqref{eq:jointposterior} for exposition, but we note that our code, implemented in \textsc{CMBLensing.jl},\footnote{\url{https://github.com/marius311/CMBLensing.jl}} calculates it automatically using a probablistic programming language given only the forward model in Eqn.~\eqref{eq:fw_model}, and calculates posterior gradients using automatic differentiation \citep{millea20}. This highlights the strength of the Bayesian approach, which allows forward models to be quickly turned into pipelines ready to use for inference. 

\section{Results}
\label{sec:results}

\subsection{Experimental setups}
We consider a number of CMB surveys representative of the current and upcoming experimental landscape.
The experimental specifications are listed in Tab.~\ref{tab:exp_specs} (taken from \citep{raghunathan2022a,raghunathan2022b}).

Our noise spectra modeling includes a contribution from both instrumental and atmospheric noise as
\begin{equation}
    N_{\ell} = \Delta_T^2 \left[1+ \left(\frac{\ell}{\ell_{\rm knee}}\right)^{-\alpha_{\rm knee}} \right],
\end{equation}
where $\Delta_T^2$ is the detector white noise while $\ell_{\rm knee}$ and $\alpha_{\rm knee}$ parametrize the atmospheric $1/f$ noise.

We do not include emission from astrophysical foregrounds in our simulations but incorporate their impact by considering the effective noise levels after component separation through harmonic-space Internal Linear Combination (ILC) methods \citep[e.g.,][]{tegmark96,delabrouille09}. 
We model the total covariance between different frequency channels at each multipole $\ell$ as a linear mixture of CMB, foregrounds, and noise contributions
\begin{equation}
    \mathbf{C}_{\ell} = \mathbf{e}\mathbf{e}^\dagger C_{\ell}^{\rm CMB} + \mathbf{C}_{\ell}^{\rm FG} + \mathbf{N}_{\ell},
\end{equation}
where $\mathbf{e}$ is a column vector of all ones (for maps in thermodynamic temperature units).
The foreground contributions in intensity include emission from radio point sources, the cosmic infrared background, diffuse kinematic and thermal Sunyaev-Zel'dovich (kSZ/tSZ) signals.
The polarized CMB sky contains emissions from Galactic dust and synchrotron.
For simplicity, we adopt the "standard" ILC noise that minimizes the total variance of the final CMB+kSZ map, so that the post-component separation ILC noise curve can be calculated as:
\begin{equation}
    N^{\rm ILC}_{\ell} = \left[ \sum_{ij} \left(\mathbf{C}^{-1}\right)^{ij}_{\ell}\right]^{-1}.
\end{equation}

Our forecasts are performed on 512\,{$\times$}\,512 pixel flat-sky maps with 2\,arcmin pixels, covering roughly 300\,deg$^2$ of sky. These are then rescaled by an $f_{\rm sky}$ factor to match the expected sky coverage of each survey. Because both lensing and patchy screening are spatially local operations, we expect this scaling will give results very similar to having performed the forecasts on larger sky areas directly, but at much lower computational cost. We do not include any masking, which we expect will be a ${<}\,10\%$ effect \cite{millea22}.

\begin{table*}[]
\begin{tabular}{c|c|cccccc|cccccc}
\hline
 \multirow{2}{*}{Experiment}         & \multirow{2}{*}{$f_{\rm sky}$} & \multicolumn{6}{c|}{Beam $\theta_{\rm FWHM}$ [arcmin]}                                                                                                               & \multicolumn{6}{c}{Noise $\Delta_T [\mu$K-arcmin]}                                                                           \\ \cline{3-14} 
                                    &                                & 30                   & 40                   & 90                   & 150                  & 220                  & 270                  & 30                 & 40                 & 90         & 150        & 220        & 270                \\ \hline
\multirow{3}{*}{SPT-3G}             & \multirow{3}{*}{3.6\%}         & \multirow{3}{*}{-}   & \multirow{3}{*}{-}   & \multirow{3}{*}{1.7} & \multirow{3}{*}{1.2} & \multirow{3}{*}{1.0} & \multirow{3}{*}{-}   & \multirow{3}{*}{-} & \multirow{3}{*}{-} & 3.0        & 2.2        & 8.8        & \multirow{3}{*}{-} \\
                                    &                                &                      &                      &                      &                      &                      &                      &                    &                    & (1200,3.0) & (1200,4.0) & (1200,4.0) &                    \\
                                    &                                &                      &                      &                      &                      &                      &                      &                    &                    & \multicolumn{3}{c}{(300,1)}          &                    \\ \hline
\multirow{3}{*}{SO} & \multirow{3}{*}{40\%}          & \multirow{3}{*}{7.4} & \multirow{3}{*}{5.1} & \multirow{3}{*}{2.2} & \multirow{3}{*}{1.4} & \multirow{3}{*}{1.0} & \multirow{3}{*}{0.9} & 71.0               & 36.0               & 8.0        & 10.0       & 22.0       & 54.0               \\
                                    &                                &                      &                      &                      &                      &                      &                      & \multicolumn{6}{c}{(1000,3.5)}                                                                      \\
                                    &                                &                      &                      &                      &                      &                      &                      & \multicolumn{6}{c}{(700,1.4)}                                                                       \\ \hline
\multirow{3}{*}{S4-wide}            & \multirow{3}{*}{67\%}          & \multirow{3}{*}{7.3} & \multirow{3}{*}{5.5} & \multirow{3}{*}{2.3} & \multirow{3}{*}{1.5} & \multirow{3}{*}{1.0} & \multirow{3}{*}{0.8} & 21.8               & 12.4               & 2.0        & 2.0        & 6.9        & 16.7               \\
                                    &                                &                      &                      &                      &                      &                      &                      & (471,3.5)          & (428,3.5)          & (2154,3.5) & (4364,3.5) & (7334,3.5) & (7308,3.5)         \\
                                    &                                &                      &                      &                      &                      &                      &                      & \multicolumn{6}{c}{(700,1.4)}                                                                       \\ \hline
\multirow{3}{*}{S4-deep}            & \multirow{3}{*}{3\%}           & \multirow{3}{*}{8.4} & \multirow{3}{*}{5.8} & \multirow{3}{*}{2.5} & \multirow{3}{*}{1.6} & \multirow{3}{*}{1.1} & \multirow{3}{*}{1.0} & 4.6                & 2.94               & 0.45       & 0.41       & 1.29       & 3.07               \\
                                    &                                &                      &                      &                      &                      &                      &                      & (1200,4.2)         & (1200,4.2)         & (1200,4.2) & (1900,4.1) & (2100,3.9) & (2100,3.9)         \\
                                    &                                &                      &                      &                      &                      &                      &                      & (150,2.7)          & (150,2.7)          & (150,2.6)  & (200,2.2)  & (200,2.2)  & (200,2.2)          \\ \hline
\end{tabular}
\caption{Technical specifications for the CMB surveys considered in this work. We show the frequency band centers (in GHz), FWHM apertures (in arcmin),  temperature  white-noise levels $\Delta_T [\mu$K-arcmin] (the polarization white-noise level is $\Delta_P=\sqrt{2}\Delta_T$), and atmospheric noise parameters (middle and bottom rows in each subpanel show the numbers for temperature and polarization, respectively).}
\label{tab:exp_specs}
\end{table*}

\subsection{Joint MAPs}
As a first check, we compute the joint maximum a posteriori (MAP) estimate of the gravitational potential, $\phi$, and the optical depth, $\tau$, on a suite of simulated data. MAP estimates are only an intermediate step towards forecasting or inference since they have biased power spectra, but they are useful as a sanity check and to build intuition about the problem. They are computed by fixing the spectral amplitudes to some fiducial value (here unity) and maximizing the posterior probability
\begin{equation}
    (\hat{f}_J,\hat{\phi}_J,\hat{\tau}_J) \;=\; \argmax_{f,\phi,\tau} \; \log \mathcal{P}(f,\phi,\tau \,|\, \theta, d).
\end{equation}
We iteratively maximize the posterior using the coordinate descent algorithm introduced in \cite{millea19}, with a modified Newton-Raphson step that performs joint $(\phi,\tau)$ updates.

\begin{figure}
\includegraphics[width=\columnwidth]{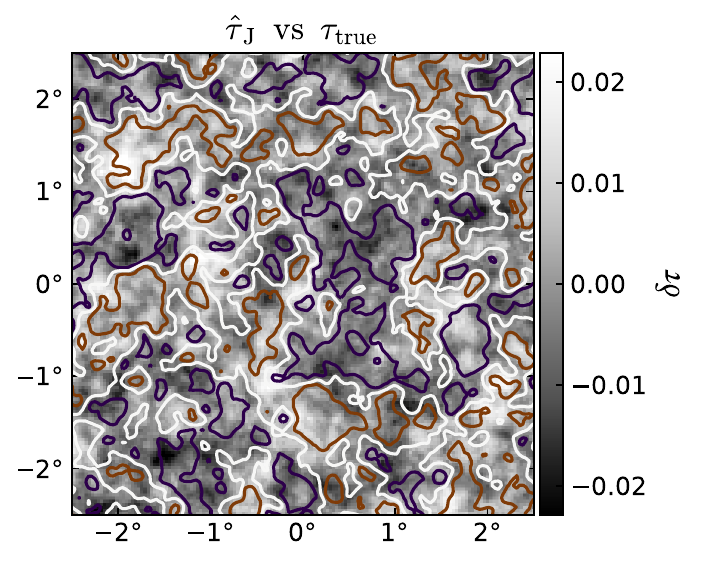}
\caption{An example maximum a posteriori (MAP) optical depth reconstruction, $\hat\tau_{\rm J}$, shown as contours, compared to the true field for the input simulation, $\tau_{\rm true}$, shown as the grayscale map. We assume S4-deep noise levels, and here for clarity show just a small cutout of the larger maps which are reconstructed as part of the forecasting procedure. Some partial correlation is visible by eye.}
\label{fig:MAP}
\end{figure}

\begin{figure}
\includegraphics[width=\columnwidth]{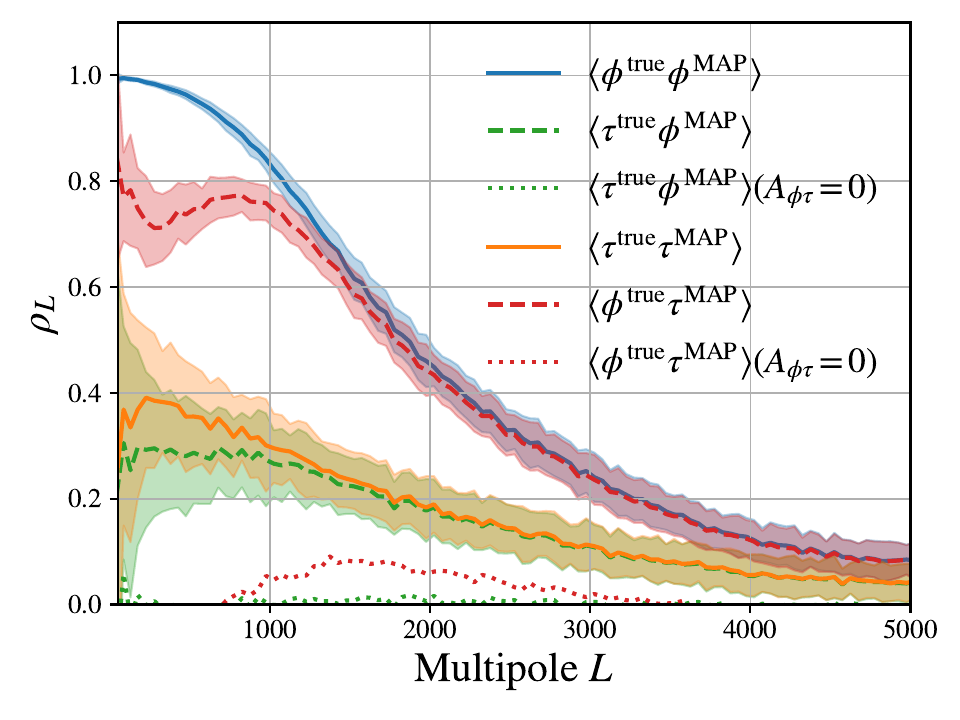}
\caption{Cross-correlation coefficient between the input CMB lensing $\phi$ and optical depth $\tau$ fields and the corresponding maximum a posteriori (MAP) estimates from simulated data, assuming S4-deep noise levels. The solid line represents the mean correlation coefficient across $N_{\rm sim}=100$ simulations while the shaded areas denote the $1\sigma$ scatter for a single realization. The average cross-correlation coefficient between the reconstructed MAP $\tau$ and input $\phi$ maps is shown as a dashed red line (the opposite case is denoted by the dashed green line). For reference, we also include the correlation coefficient between the reconstructed MAP and true maps when the input $\tau$ and $\phi$ fields are not intrinsically correlated, i.e. $A_{\phi\tau}=0$, as the dotted lines (same color coding as the $A_{\phi\tau}=1$ case).}
\label{fig:rho_MAP}
\end{figure}

Fig.~\ref{fig:MAP} shows a typical $\hat\tau_{\rm J}$ as compared to the simulation truth, assuming S4-deep noise levels and using only polarization. The contour levels are set at 0 and $\pm1$ times the pixel RMS of the MAP reconstruction, and can be visibly seen to trace the true field, with regions of low $\tau$ encircled by blue contours regions of high $\tau$ encircled by orange contours. 
More quantitatively, in Fig.~\ref{fig:rho_MAP}, we show the cross-correlation coefficient $\rho_{\ell}$ at different scales between the true $(\phi,\tau)$ maps and the corresponding joint MAP estimates from the simulated data, averaged over several realizations. We see that the MAP $\phi$ reconstruction is correlated to the true $\phi$ map, as expected, and that the MAP $\tau$ reconstruction is correlated to the true $\tau$, indicating the potential for a detection. We also see that the MAP $\tau$ is correlated to the true $\phi$. While this is also expected to some degree, given that the true $\tau$ and $\phi$ fields are intrinsically correlated, the fact that $\langle\tau^{\rm true}\phi^{\rm MAP}\rangle \neq \langle\tau^{\rm MAP}\phi^{\rm true}\rangle$ implies an additional lensing-induced bias in the power spectra of the $\tau$ MAPs (beyond the suppression of power which is typical of MAP estimates). This is not surprising, as it is known that the $\tau$ QE contains a lensing-induced bias \cite{su11,namikawa21}, and MAP estimates asymptote to the QE in the high-noise limit. We note that our Bayesian approach implicitly deals with this when marginalizing over the $\{f,\phi,\tau\}$ maps, allowing unbiased inference of parameters which control the theory auto and cross-correlation spectra without having to explicitly compute any such bias. As an additional check, in Fig.~\ref{fig:rho_MAP} we include the correlation coefficients between the true $\tau$ and MAP $\phi$ (and vice versa) when we turn off the intrinsic correlation between the true $\tau$ and $\phi$ fields.

\subsection{Forecasting results}

\begin{table}
\begin{center}
\begin{tabular}{ll|c|c|c}
 & $\ell_{\rm max}$ & $100\,\sigma(A_{\phi\phi})$ & $\sigma(A_{\tau\tau})$ & $\sigma(A_{\phi\tau})$  \\ 
\toprule
SO      & 3000 & 1.5\,{--}\, & 2.3\,{--}\, & 0.51\,{--}\, \\
        & 5000 & 1.5\,{--}\, & 2.2\,{--}\, & 0.51\,{--}\, \\
\midrule         
SPT-3G  & 3000 & 1.4\,{--}\,1.4 & 0.97\,{--}\,3 & 0.29\,{--}\,0.44 \\
        & 5000 & 1.4\,{--}\,1.4 & 0.95\,{--}\,3 & 0.28\,{--}\,0.43 \\
\midrule         
S4-wide & 3000 & 0.38\,{--}\,0.38 & 0.29\,{--}\,2.6 & 0.080\,{--}\,0.18 \\
        & 5000 & 0.37\,{--}\,0.41 & 0.29\,{--}\,2.6 & 0.078\,{--}\,0.16 \\
\midrule         
S4-deep & 3000 & 0.57\,{--}\,0.55 & 0.13\,{--}\,1.3 & 0.066\,{--}\,0.110 \\
        & 5000 & 0.55\,{--}\,0.51 & 0.12\,{--}\,1.3 & 0.060\,{--}\,0.088 \\
\midrule         
S4      & 3000 & 0.31\,{--}\,0.31 & 0.12\,{--}\,1.2 & 0.050\,{--}\,0.088 \\
        & 5000 & 0.31\,{--}\,0.30 & 0.11\,{--}\,1.2 & 0.047\,{--}\,0.072 \\
\bottomrule        
\end{tabular}
\end{center}
\caption{Forecasted 1\,$\sigma$ uncertainties from polarization for various configurations. Each entry gives a range corresponding to pessimistic and optimistic models (except in the SO case where only the optimistic model is considered). In all cases, the amplitude parameters are defined such that their fiducial value is unity (and note that the first column is multiplied by 100).}
\label{table:results}
\end{table}

The key challenge for inference or forecasting is the marginalization over the $\{f,\phi,\tau\}$ fields. While we could use an exact sampling algorithm like Hamiltonian Monte Carlo (HMC) to perform this marginalization, we instead use the approximate but much faster Marginal Unbiased Score Expansion (MUSE) method \cite{seljak17,millea22,millea22b}. For data vectors as large as ours here, MUSE is unbiased, but in theory can be sub-optimal. However, it was demonstrated in \cite{millea22} that for gravitational lensing inference, it achieves uncertainties which are ${<}\,10\%$ away from the optimal Fisher limit. We do not expect the addition of patchy screening to qualitatively change this behavior, motivating its use here. Although we expect our results are close to optimal, they should be regarded as an upper-bound on possible uncertainties, and whether tighter constraints are possible could be checked by analyzing our posterior probability function with HMC.

Following \cite{millea22}, we use MUSE to compute the $J$ and $H$ matrices for the parameters set $\{A_{\phi\phi}, A_{\tau\tau}, A_{\phi\tau}\}$, then use these to form the marginal posterior covariance matrix. The $J$ matrix is the covariance of gradients of the log posterior with respect to parameters at the joint MAP, and the $H$ matrix is akin to a ``response function,'' which quantifies how the gradient with respect to each parameter responds to injecting a shift in another parameter into the simulated data. In practice, the computation involves only obtaining MAP estimates of the kind pictured in Fig.~\ref{fig:MAP} for a suite of data simulations, and evaluating gradients of the posterior in Eqn.~\eqref{eq:jointposterior} at these points in parameter space. The MUSE covariance, much like the inverse Fisher matrix, is already averaged over data realizations, and so directly gives us our desired forecast. 

\begin{figure}
\includegraphics[width=\columnwidth]{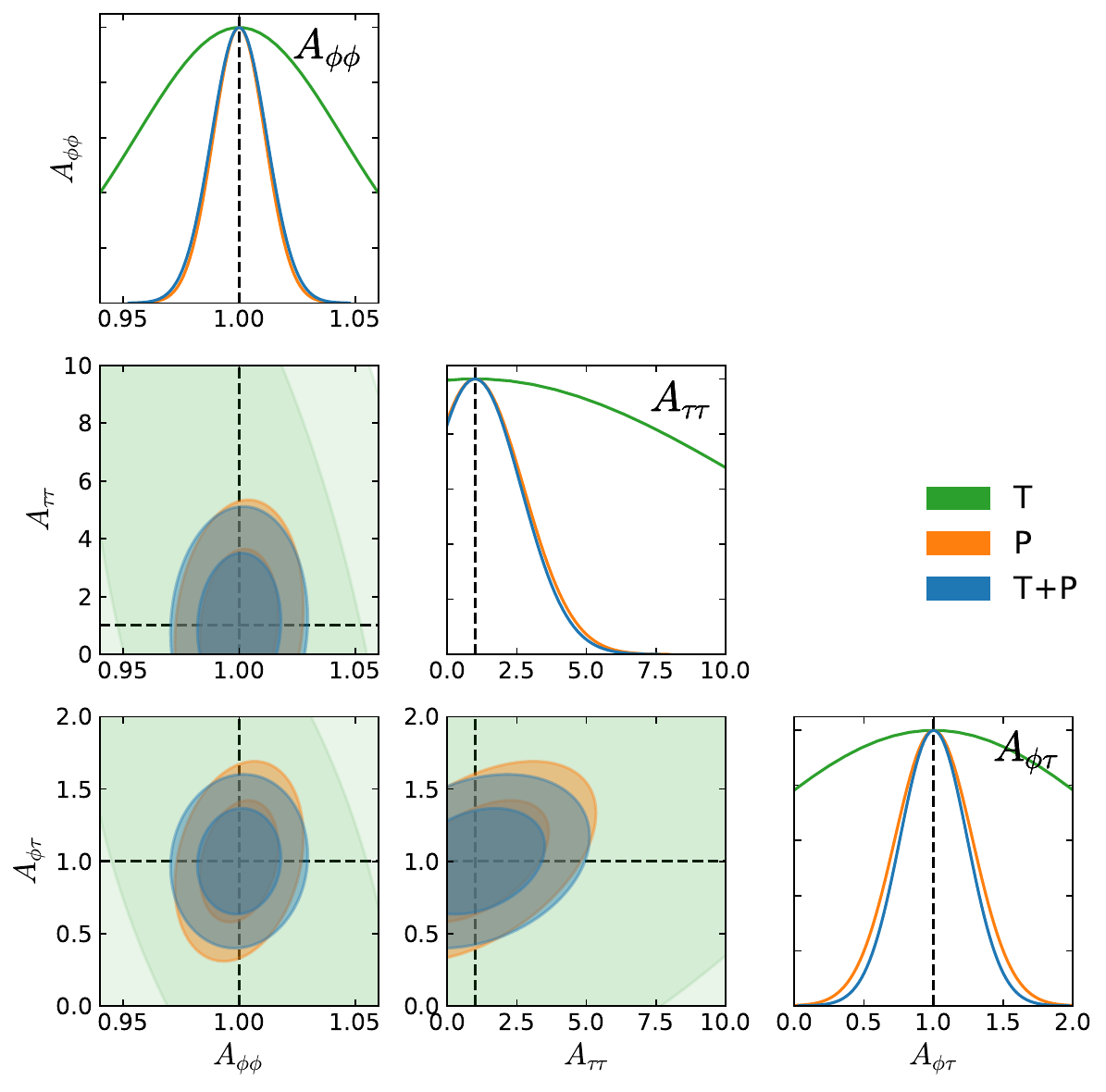}
\caption{Forecasted constraints on lensing and screening amplitudes from temperature and/or polarization for SPT-3G, assuming $\ell_{\rm max}\,{=}\,3000$. We find that at these noise-levels (and below), temperature adds negligible information in terms of the overall amplitude parameters.}
\label{fig:T_vs_P}
\end{figure}

We begin by examining the information content in temperature vs. polarization. In Fig.~\ref{fig:T_vs_P}, we plot the forecasted covariance for the SPT-3G configuration with $\ell_{\rm max}\,{=}\,3000$ as one and two-dimensional marginalized contours. We find that there is almost no weight from temperature on constraints to the overall lensing or screening amplitudes. It is already known that at SPT-3G noise levels, the total lensing $S/N$ is highly dominated by polarization, and we find the same is true for the screening effect. We expect this will only get more drastic as noise levels decrease, thus for the forecasts described below, we consider only polarization for simplicity. By doing so, however, we are underestimating constraints possible with SO, which has higher noise-levels and could receive a non-negligible contribution from temperature. Doing such a forecast would depend on more complex assumptions about temperature foregrounds, so we leave this to a future work.

We now consider the potential for detecting the screening signal from different ongoing and future CMB surveys. These results are summarized in Table~\ref{table:results}. Below, we will use the term ``detection'' to refer to constraining a non-zero amplitude parameter at $3\,\sigma$. With SPT-3G, it will be possible to make a detection of the patchy screening correlation with the lensing potential given the optimistic model, but not given the pessimistic model. With SO polarization, it is not possible to detect either model. In general, detecting the auto-correlation of the patchy screening is comparatively harder than detecting the cross-spectrum, and is not possible with SPT-3G or SO regardless of model. For CMB-S4, either scenario leads to a guaranteed detection of the cross-correlation, and additionally, the auto-correlation is detectable given the optimistic model. We note that for CMB-S4, the deep survey contributes slightly more weight than the wide survey. With the pessimistic model, a guaranteed detection of the auto-correlation of the screening effect will require at least a 5th generation CMB survey.

\subsection{Comparisons with other works}

Forecasts similar to ours have been computed in previous works, based on the QE instead of on Bayesian inference. We do not attempt a careful quantitative comparison with other works, but comment briefly on similarities. \citet{roy18} computed the total $S/N$ in the QE $\tau$ auto-spectrum, while arguing that lensing-induced biases were ignorable for their purposes due to sufficient $S/N$. One of their forecasts considers a reionization model with $\bar R\,{=}\,5$ and $\sigma_{\ln\!R}\,{=}\,\ln(2)$, similar to our optimistic model, with a similar total optical depth and lower assumed noise levels for S4. For this case, they predict a $5\,\sigma$ detection is possible with S4. Our S4 optimistic forecast gives a $10\,\sigma$ detection of $A_{\tau\tau}$, despite our higher noise assumptions. \citet{feng19} perform QE forecasts which account for lensing-induced biases, and additionally consider the cross-spectrum between lensing and screening fields. We have computed theory spectra using the reionization parameters quoted in their paper, visually verifying similar $C_\ell^{\phi\tau}$ spectra, and performed a forecast for S4 with their noise assumptions (more optimistic than ours). Their result gives a total $S/N$ of $22\,\sigma$ in the $\phi(EB)\tau(EB)$ estimator, and we find a $23\,\sigma$ detection of non-zero $A_{\phi\tau}$ for this case. 

We also note that amplitude parameters only capture the total $S/N$, and for spectra which are very red at small scales such as the lensing and patchy screening spectra, the $S/N$ is mainly limited by the large-scale sample variance, not the reconstruction noise. This is the reason why we do not see more improvement with $\ell_{\rm max}$ in Tab.~\ref{table:results}, and is also a reason why the above comparison is an incomplete test of the exact improvements upon the QE which are happening. An internal scale-dependent comparison of Bayesian and QE estimates of patchy screening, however, is outside the scope of this work, as forming an unbiased QE patchy reconstruction pipeline akin to \cite{feng19} is significantly non-trivial, and regardless, we expect theoretically that the Bayesian pipeline will extract the same or more information than the QE pipeline. Our rough comparisons above indeed appear to be consistent with this expectation.

\begin{figure}
\includegraphics[width=\columnwidth]{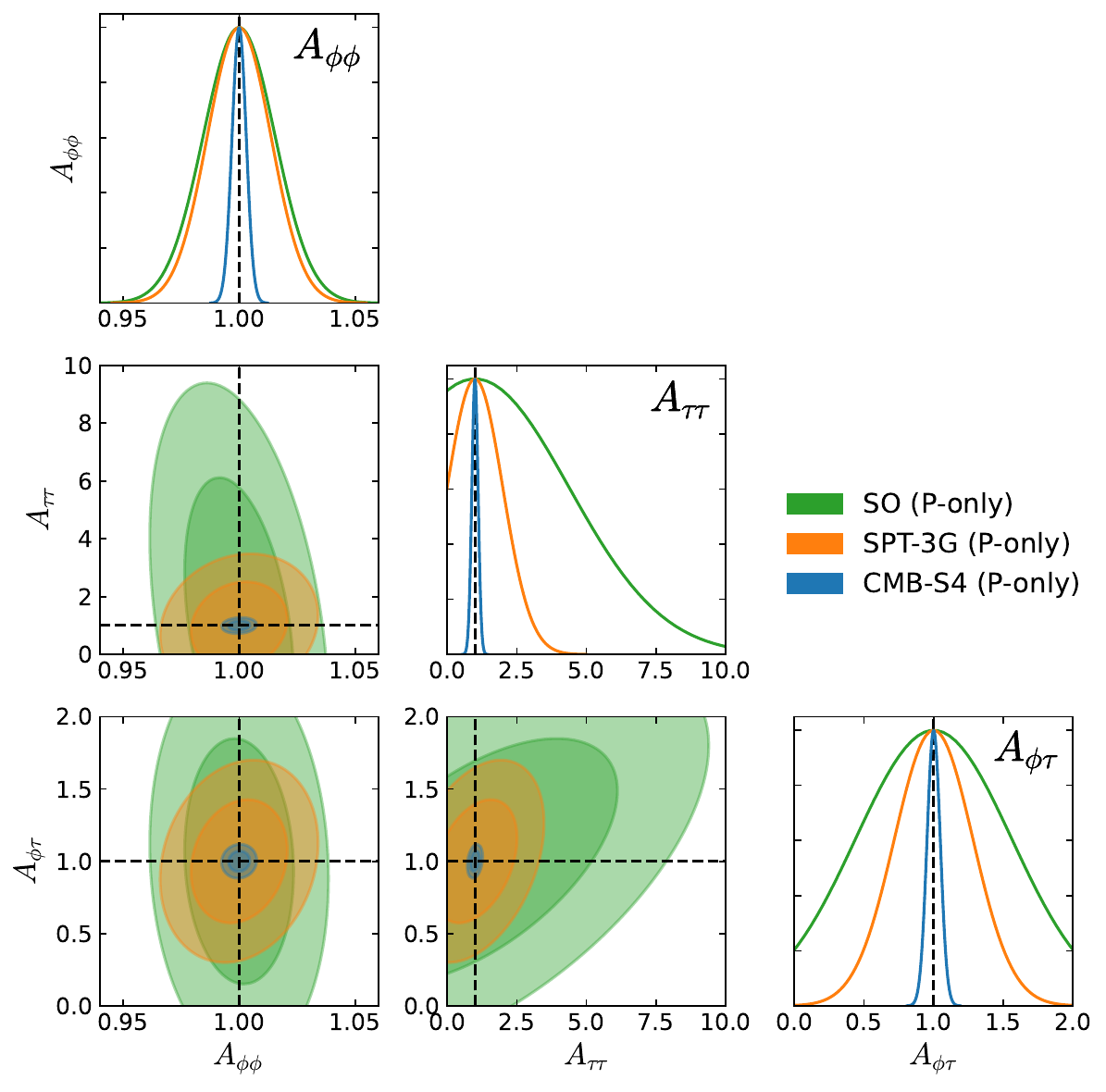}
\caption{Forecasted constraints from polarization for different experiments, given the optimistic reionization model and assuming $\ell_{\rm max}\,{=}\,5000$.}
\label{fig:S4_vs_3G}
\end{figure}

\section{\label{sec:conclusions} Conclusions}
In this work, we have developed a Bayesian model which can be used to simultaneously extract fluctuations in the optical depth, $\tau$, and gravitational lensing potential, $\phi$, together with delensed and descreened CMB fields, as well as the theory spectra which describe their statistics. Here, we have used MUSE to infer the (amplitudes of) the theory spectra and to forecast their uncertainties from upcoming probes. In the future, this code can directly be used to  directly obtain estimates of the amplitudes from data, or can be used with HMC sampling or other inference tools to obtain joint inference on the spectra and fields together.

The Bayesian approach allows us to optimally recover all available information from the data, and implicitly handles any bias subtraction, such as the lensing-induced bias of the $\tau$ QE. In practice, we find its strength is its simplicity, requiring us only to code up a field-level forward model which includes lensing and screening, with the posterior probability function and its gradients automatically derived, and MUSE providing quick forecasting which requires only solving optimization problems. This type of analysis can readily be extended to include other secondary signals or foregrounds, and its equivalent could be applied to other problems in cosmology which have computable forward models. The trade-off is that any real analysis requires high-fidelity modeling of observed data, but these types of pipelines are being developed, including for the next generation of CMB experiments. 

We have found that at the low-noise levels characteristic of the current most sensitive and upcoming CMB surveys ($\Delta_P \lesssim 5\,\mu$K-arcmin), polarization data carry the majority of the constraining power on patchy reionization. 
Owing to the faintness of the patchy screening effect, a definitive detection of the optical depth fluctuations will prove challenging in auto-correlation.
Depending on the details of the reionization history, a next-generation experiment like CMB-S4 might be able to detect the signal but for more pessimistic scenarios we would require a futuristic Stage-5 mission.
On the other hand, measuring the cross-correlation between $\phi$ and $\tau$ holds great promise for detecting optical depth anisotropies.
For the full-depth SPT-3G survey, we forecast a 3\,$\sigma$ detection of the cross-correlation signal in the optimistic scenario, while CMB-S4 will be able to detect the signal with $S/N \gtrsim 10$ even in more pessimistic scenarios.

The next-generation of CMB surveys will map the spatial fluctuations of free electrons out to high redshifts, informing models of reionization and providing clues on the gas physics around galaxies \citep[][]{CMB-S4_22,delabrouille19}.
Moving forward, a compelling avenue to explore would be to extend our modeling to provide direct constraints on phenomenological or physical reionization parameters, or the joint analysis with observables probing other phases of matter,
such as Compton-$y$ or 21-cm maps \citep[e.g.,][]{meerburg13,namikawa21,roy21}, or the galaxy distribution itself.

\begin{acknowledgments}
We thank Gil Holder and Emmanuel Schaan for useful discussions and Srinivasan Raghunathan for advice on foreground cleaning.
Part of this research used resources of the National Energy Research Scientific Computing Center, which is supported by the Office of Science of the U.S. Department of Energy under Contract No. DE-AC02-05CH11231.
\end{acknowledgments}

\appendix

\section{An halo model for reionization} \label{app:halo_model}
In this appendix, we briefly review the basics of the analytic halo model formalism we use to predict the power spectrum of the optical depth field ($\tau\tau$) and its cross-correlation with CMB lensing potential ($\phi\tau$). 

The model presented here is largely based on the phenomenological parametrization introduced in \citep{wang06,mortonson07,dvorkin09}.
In this picture, dark matter halos host the \textsc{Hii} regions surrounding the ionizing sources such as galaxies or quasars.
These regions are modeled as fully ionized spherical bubbles of radius $R$ inside the neutral intergalactic medium.
As time progresses, the ionized bubbles grow and eventually overlap with each other until the Universe is fully reionized.

To calculate the optical depth auto-spectrum $C_\ell^{\tau\tau}$ we need a model for the three-dimensional power spectrum of the density-weighted ionization fraction fluctuations, $P_{\delta_e \delta_e}(k,z)$. 
This quantity captures the physics and morphology of reionization which sensitively depend on the size distribution and evolution of the ionized bubbles, which is the first ingredient of the model.

Assuming that the bubble sizes are drawn from a log-normal distribution with characteristic size $\bar{R}$ and variance given by $\sigma_{\ln\!R}^2$ \citep[e.g.,][]{zahn06,friedrich11,lin16}, we can write
\begin{equation}
    \label{eq:bubble_dist}
    P(R) = \frac{1}{R} \frac{1}{\sqrt{2\pi\sigma_{\ln\!R}^2}} e^{-\left[\ln (R/\bar{R})\right]^2/(2\sigma_{\ln\!R}^2)},
\end{equation}
so that the average bubble volume can be calculated as
\begin{equation}
    \langle V_b\rangle \equiv \int \mathrm{d} R V_b(R) P(R) = \frac{4 \pi \bar{R}^3}{3} e^{9 \sigma_{\ln\!R}^2 / 2}.
\end{equation}
While in principle both $\bar{R}$ and $\sigma_{\ln\!R}^2$ are function of $z$, and would require complex radiative transfer simulations to be evaluated, we follow the previous works and assume that they do not evolve over cosmic time.

The model further assumes that the probability that a given point in space $\mathbf{r}$ is ionized is determined by a Poisson process with fluctuating mean,
\begin{equation}
    \left\langle x_e(\mathbf{r})\right\rangle_{P}=1-e^{-n_b(\mathbf{r}) \langle V_b\rangle},
\end{equation}
where $n_b(\mathbf{r})$ is the density of \textsc{Hii} bubbles and the brackets $\langle\dots\rangle_P$ denote averaging over the Poisson process. 

The \textsc{Hii} bubbles are then assumed to trace the large-scale structure with some bias $b$,
\begin{equation}
    n_b(\mathbf{r}) = \bar{n}_b \left(1+b\delta_W(\mathbf{r})\right),
\end{equation}
where $\delta_W=\int d^3\mathbf{r'} \delta(\mathbf{r}) W_R(\mathbf{r}-\mathbf{r'})$ is the matter overdensity $\delta$ smoothed by a top-hat window of radius $R$ which in Fourier space is given by:
\begin{equation}
    W_R(k)=\frac{3}{(k R)^3}[\sin (k R)-k R \cos (k R)].
\end{equation}
For simplicity, we assume that the bubble bias does not evolve with redshift and is independent of the bubble size. 
Following previous works \citep[e.g.,][]{dvorkin09,roy18}, in this paper we fix $b=6$.
With these definitions, the average bubble number density can be related to the mean ionization fraction as $\bar{n}_b=- \ln{(1-\bar{x}_e)}/\langle V_b\rangle$. 
For completeness, we further define the two-bubble and one-bubble window functions averaged over the bubble radius distributions respectively as:
\begin{align}
\langle W_R \rangle(k) &= \frac{1}{\langle V_b\rangle} \int_0^{\infty} {\rm d}R\,P(R)V_b(R)W_R(kR) \\
\langle W^2_R \rangle(k) &= \frac{1}{\langle V_b\rangle^2} \int_0^{\infty} {\rm d}R\,P(R)\left[V_b(R)W_R(kR)\right]^2. 
\end{align}

In analogy with the standard halo model of large-scale structure \citep[e.g.,][]{cooray02}, we can decompose the \textsc{Hii} power spectrum into the sum of a 1-bubble term, which dominates at scales $r \ll R$, and a 2-bubble contribution that prevails at larger scales, $r \gg R$:  $P_{\delta_{e} \delta_{e}}(k) = P_{\delta_{e} \delta_{e}}^{1b}(k)+P_{{\delta_{e} \delta_{e}}}^{2b}(k)$.\footnote{Here we suppress the redshift dependence for brevity.}
The analytic expressions of the 2-bubble and 1-bubble terms of the ionized hydrogen power spectrum are given respectively by
\begin{align}
P_{\delta_{e} \delta_{e}}^{2b}(k) &=\left[\left(1-\bar{x}_{e}\right) \ln \left(1-\bar{x}_{e}\right) b\left\langle W_{R}\right\rangle(k) - \bar{x}_e\right]^{2} P_{\delta \delta}(k) \\
P_{\delta_{e} \delta_{e}}^{1b}(k) &=\bar{x}_{e}\left(1-\bar{x}_{e}\right)\left[\left\langle V_{b}\right\rangle\left\langle W_{R}^{2}\right\rangle(k)+ \Tilde{P}(k) \right], 
\end{align}
where $\Tilde{P}(k)$ can be approximated as \citep[][]{wang06,mortonson07}
\begin{equation}
    \Tilde{P}(k) \approx \frac{P_{\delta \delta}(k)\left\langle V_{b}\right\rangle\left\langle\sigma_{R}^{2}\right\rangle}{\sqrt{P_{\delta \delta}^{2}(k)+\langle V_{b}\rangle^2\langle\sigma_{R}^{2}\rangle^{2}}},
\end{equation}
and $\langle\sigma_{R}^{2}\rangle$ is the matter density variance top-hat smoothed over the bubble radius distribution.
In the formulae above, $P_{\delta\delta}(k)$ denotes the linear matter power spectrum.

Similarly, the 2- and 1-bubble terms relative to the cross-power spectrum between the ionization fraction fluctuations and the matter density contrast can be evaluated as
\begin{align}
P_{\delta\delta_{e}}^{2b}&=\left[\bar{x}_{e}-\left(1-\bar{x}_{e}\right) \ln \left(1-\bar{x}_{e}\right) b\left\langle W_{R}\right\rangle(k)\right] P_{\delta \delta}(k)\\
P_{\delta\delta_{e}}^{1 b} &\approx 0,
\end{align}
where the term $P_{\delta\delta_{e}}^{1 b}$ is zero since we assume the bubbles are completely ionized.

\nocite{*}
\bibliographystyle{aasjournal}
\bibliography{apssamp}

\end{document}